\documentstyle[aps,prb,epsf]{revtex}
\begin{document}
\title{\Large \bf ELECTRON-ENERGY-LOSS SPECTROSCOPY OF THE ${\bf C_{60}}$ MOLECULE}
\author{{\bf D.A. Gorokhov\footnote{present address: Theoretische Physik,
ETH-H\"onggerberg, CH-8093, Z\"urich, Switzerland; 
e-mail: gorokhov@itp.phys.ethz.ch},
 R.A. Suris,
 V.V. Cheianov\footnote{present address: Institute for Theoretical Physics,
Uppsala University, Box 803, S-75108, Uppsala, Sweden; 
e-mail: sheianov@helene.teorfys.uu.se}}
\\{\it Ioffe Physical Technical Institute}
\\{\it Politehnicheskaya str., 26, 194021, S.-Petersburg, Russia}}
\maketitle
\begin{abstract}
In this work we present a theoretical study of EELS
(electron-energy-loss spectroscopy) experiments on the ${\rm C}_{60}$ molecule.
Our treatment of the problem is based on the simple two-fluid model
originally proposed for the description of plasma oscillations
in graphite and fullerenes (see Refs.\cite{14,15,16}). It is shown that
in spite of the simplicity of the model
 the calculated intensities of the EELS peaks are in
good agreement with experimental data which may indicate
that the model can be used as a simple and effective tool for the investigation
of the collective behaviour of electrons in fullerene systems.
\end{abstract}

\ \ \ \ \ \ \ \ \ \ \ \ \ {Phys. Lett. A {\bf 223}, 116 (1996)}

\section{Introduction}

Investigations of plasmons in small 
clusters of nanometer size range have a long history.
Originally plasma oscillations were studied in small ($R=2\div 10$ nm)
metal particles. 
It turned out that
plasmons in metal clusters can be well described by linearized
hydrodynamic equations \cite{1,2,3}.
The positions of plasmon peaks, their number and fast electron cross-sections
calculated within these equations
are in good agreement with experimental data.
New possibilities of the investigation of plasmons in small objects
appeared after the discovery of fullerenes and the cheap methods of their
synthesis. Plasmon oscillations were observed in fullerene molecules \cite{4,5,6}, 
carbon nanometer size tubes \cite{7} and
multishell fullerenes \cite{8}.

An important feature of fullerenes is the fact that these
molecules are more or less of a spherical form and their electrons
can be considered to be confined to the surface of the sphere\cite{9}.
Hence, it seems to be a good starting point to consider the ${\rm C}_{60}$
molecule as a rolled up graphite plane. This approximation is expected
to work well for the description of collective excitations.

The graphite plane consists of regular hexagons formed by carbon atoms.
Each carbon atom is connected by 3 $\sigma$-bonds with its three
neighbours, meanwhile $\pi$-electrons form a common $\pi$-system
consisiting of orbitals perpendicular to the plane. 
In linear response theory
$\pi$ and
 $\sigma$-electron subsystems 
can be considered to be independent with respect to the
electrical field parallel to the graphite plane because dipole
transitions from $\pi$ to $\sigma$-orbitals are prohibited.
This means that we can describe the plasmons in graphite whose
wave vector is parallel to the plane as the plasmons in a ``two-liquid"
electron system;
it is proposed that both the liquids move in a common
average potential.

The graphite loss function -Im$(1/{\epsilon_{\bot}(\omega)})$
with respect to the external electric field perpendicular
to the $c$-axis shows two peaks at 7 and 28 eV 
(see Refs.\cite{10,11,12,13}).
They are called $\pi$- and $\sigma$- (or $``\pi +\sigma "$-) plasmons.
 Strictly speaking such a division is rather conventional.
In fact, both
$\pi$- and $\sigma$-electrons make a nonzero contribution to
$\sigma$- and $\pi$-plasmons respectively.

In the articles\cite{14,15} Cazaux used the two-fluid
model in order to explain the results of EELS studies on graphite.
The same idea was used in Ref.\cite{16} for the description
of plasmons in the ${\rm C}_{60}$ molecule:
it was proposed that $\pi$-electrons
be affected only by induced electric potential, meanwhile $\sigma$-electrons
be governed also by a restoring force.
This force appears in the model via an effective description
of a contribution of valent electrons to the polarizability
\cite{17,18,19}. 

Using the two-fluid model Barton and Eberlein predicted the existence
of plasma oscillations in the ${\rm C}_{60}$ molecule which can be classified
by the 
spherical quantum numbers $l$ and $m$ and 
the index $j$ ($j=1,2$) which corresponds to
$\pi$- and $\sigma$-plasmons \cite{16}. 
However, the plasmons with large $l$
cannot exist because of their strong damping.

EELS studies on ${\rm C}_{60}$ 
(see Refs.\cite{4,5}) show a peak at 6.5-7 eV (also called
``$\pi$-plasmon") and a rather broad ``$\sigma$-plasmon" peak at 15-16 eV .
 The former peak is identified usually as the $l=1$ $\pi$-plasmon,
meanwhile the latter peak is considered to be a combination of the
$l=1$ $\sigma$-plasmon and probably the plasmons with $l>1$. However the
existence of multipole plasmons in ${\rm C}_{60}$ is still not clear.
In Ref.\cite{6} a small peak at 28 eV was observed and identified
as the $l=1$ $\sigma$-plasmon.

The main goal of this work is to calculate the cross-section
of fast electrons scattered by a ${\rm C}_{60}$ molecule
and to show that the contribution of the plasmons with $l=2$
to the cross-section is enough in order to be observed in EELS
experiments. In the model we use it is possible to separate
contributions of different plasmons, i.e. one can obtain relative
cross-sections.
Using this approach we can not calculate plasmon damping.
Widths of plasmon peaks will be taken from experimental data.

\section{Quantization of plasma oscillations on a sphere
in the two-fluid model}

We shall consider electrons in the ${\rm C}_{60}$ molecule to be distributed
over the surface of a uniformly charged sphere --- the positive molecular core
in the model. The Fourier transformation of the charge density
$\delta n(\omega )$ induced by the potential $\Phi$ including both
the external and induced potentials is given by the expression
\begin{equation}
\delta n(\omega )=
-\sum_{\alpha}
\frac{e^{2}n}{m}\frac{f_{\alpha}}{{\omega _{\alpha}}^{2}-{\omega}^{2}}
\Delta_{\|} \Phi (\omega)
\label{1}
\end{equation}
where $n$ is the surface $\pi$- and $\sigma$-{\rm electron} density,
$f_{\alpha}$ and ${\omega _{\alpha}}$ are respectively the oscillator strength
and electron energy levels. The index $\alpha$ numerates the electron states.
$\Delta_{\|}$ denotes the angular part of the Laplas operator:
in the model we use
electrons cannot move perpendicularly to the surface.

As the $\pi$-electron subsystem of graphite is semimetal, we can use
the following approximation for the $\pi$-electron contribution to the
induced charge density:
\begin{equation}
\delta n_{\pi}(\omega )=
\frac{e^{2}N}{4\pi m R^{2}}\frac{1}{{\omega}^{2}}
\Delta_{\|} \Phi (\omega).
\label{2}
\end{equation}

We suggested $\omega_{\alpha}$ in equation $(\ref{2})$ to be equal to zero
and used the sum rule $\sum_{\alpha}f_{\alpha\pi}=1/4$ for the oscillator
strength of $\pi$-{\rm electrons} in $C_{60}.$ $N=60$ and
$R=0.4\ {\rm nm}$
are the number of $\pi$-electrons and the effective radius of the
sphere respectively.

On the other hand, the $\sigma$-{\rm electron} subsystem of graphite
is dielectric with the gap of order $12\ eV$ 
(see Ref.\cite{20}).
The oscillator strength of the $\sigma$-{\rm electron} subsystem
has a maximum at $\omega_{\alpha}\simeq 16\ eV.$ Hence we can use the following
approximate expression for the $\sigma$-{\rm electron} contribution
\begin{equation}
\delta n_{\sigma}(\omega )=
\frac{3 e^{2}N}{4\pi m R^{2}}\frac{1}{{\omega}^{2}-{\omega_{\sigma}}^{2}}
\Delta_{\|} \Phi (\omega).
\label{3}
\end{equation}
Here we used the sum rule $\sum_{\alpha}f_{\alpha\sigma}=3/4$ for the
$\sigma$-{\rm electron} subsystem. $\omega_{\sigma} =16\ eV$ is the characteristic
transition frequency of $\sigma$-{\rm electrons}. $1s^{2}$~-~electrons
of carbon atoms are not taken into account in the model.

The $\Phi$-field is connected with the total induced charge density
$\delta\sigma=\delta{n}_{\pi}+\delta{n}_{\sigma}$ through the Gauss
theorem
\begin{equation}
\frac{\partial\Phi}{\partial r}{\big |}_{r=R+0}-
\frac{\partial\Phi}{\partial r}{\big |}_{r=R-0}=
4\pi\delta\sigma .
\label{gauss}
\end{equation}
It is convenient to expand the density fields in Fourier series on the
sphere
\begin{equation}
\delta n_{\pi}=\sum_{lm}a_{lm}Y_{lm}(\phi,\theta),\ \ 
\delta n_{\sigma}=\sum_{lm}b_{lm}Y_{lm}(\phi,\theta),\ \ 
\sigma=\sum_{lm}s_{lm}Y_{lm}(\phi,\theta).
\end{equation}
Then using the multipole expansion of the potential $\Phi$

\[ \Phi=\left\{ \begin{array}{r@{\quad:\quad}l}
\sum_{lm}\phi_{lm}{\big ({r}/{R}\big )}^{l}Y_{lm}(\phi,\theta)\ \ \ 
& r<R,\\ 
\ \ \sum_{lm}\phi_{lm}{\big ({R}/{r}\big )}^{l+1}Y_{lm}(\phi,\theta)
& r>R \end{array} \right. \]
we can rewrite ({\ref{gauss}}) in the following form
\begin{equation}
\phi_{lm}=-\frac{4\pi R}{2l+1}s_{lm}=-\frac{4\pi R}{2l+1}(a_{lm}+b_{lm})
\label{c7}
\end{equation}
which gives for ({\ref{2}}) and ({\ref{3}})
(see also Refs.\cite{16,24})
\begin{equation}
{\ddot a}_{lm}
=-{\Omega_{l}}^{2}
(a_{lm}+3b_{lm})
\label{13}
\end{equation}
\begin{equation}
{\ddot b}_{lm}
=-3{\Omega_{l}}^{2}
(a_{lm}+b_{lm})-{\omega_{\sigma}}^{2}b_{lm}
\label{14}
\end{equation}
where
\begin{equation}
{\Omega_{l}}^{2}={\frac{4\pi e^{2}\sigma _{0}}{m R}\frac{l(l+1)}{2l+1}},\ \ 
\sigma_{0}=\frac{N}{4\pi R^{2}}.
\end{equation}

It can be easily verified by straightforward calculation 
that equations ({\ref{13}}) and ({\ref{14}}) are the equations
of motion corresponding to the Hamiltonian 
\begin{eqnarray}
H=\frac{m\sigma_{0}}{2}\sum_{lm}{\dot{A}}_{lm}{\dot{A}}_{l,-m}+
\frac{3m\sigma_{0}}{2}\sum_{lm}{\dot{B}}_{lm}{\dot{B}}_{l,-m}+
\ \ \ \ \ \ \ \ \ \ \ \ \ \ \ \ \ \ \ \ \ \
\nonumber\\
\frac{m\sigma_{0}}{2}\sum_{lm}{\Omega_{l}}^{2}
A_{lm}A_{l,-m}+
\frac{3m\sigma_{0}}{2}\sum_{lm}(3{\Omega_{l}}^{2}+{\omega_{\sigma}}^{2})
B_{lm}B_{l,-m}+
\nonumber\\
\frac{m\sigma_{0}}{2}\sum_{lm}3{\Omega_{l}}^{2}(A_{lm}B_{l,-m}+
A_{l,-m}B_{lm})
\label{eqnarray}
\end{eqnarray}
where 
the amplitudes $A_{lm}$ and $B_{lm}$ are the $(l,m)$-components of the 
displacement fields ${\bf r}_{1}$ and ${{\bf r}_{2}}$ of the electron
liquids which are connected with $a_{lm}$ and $b_{lm}$ by the following
equations
\begin{equation}
a_{lm}=\frac{\sigma_{0}}{4}\frac{\sqrt{l(l+1)}}{R^{2}}A_{lm},\ \ 
b_{lm}=\frac{3\sigma_{0}}{4}\frac{\sqrt{l(l+1)}}{R^{2}}B_{lm}.
\label{connect}
\end{equation}

The coefficients in equation
({\ref{eqnarray}})
are chosen so as to keep it equal to the hydrodynamic
Hamiltonian 
\begin{equation}
H=\frac{m\sigma_{0}}{2}\int
{\dot {\bf r}}_{1}^{2} dS+
\frac{3m\sigma _{0}}{2}\int
{\dot{\bf r}}_{2}^{2} dS+
\frac{3m\sigma _{0}{\omega_{\sigma}}^{2}}{2}\int
{{\bf r}}_{2}^{2} dS+
\frac{1}{2}\int{\phi\delta \sigma dS}
\label{18}
\end{equation}

Hamiltonian (\ref{eqnarray}) is nondiagonal in the basis
consisting of the functions $Y_{lm}(\phi,\theta)$ because of the last term.
On the other hand, this term does not mix amplitudes
with different $l$ and $|m|$ 
because of the spherical simmetry of the model,
hence the above Hamiltonian can be
diagonalized by the diagonalization of each ``block" with a number
$lm.$

We have shown that the problem can be reduced to the diagonalization
of the Hamiltonian
\begin{equation}
\frac{m\sigma_{0}}{2}{\dot {A}}^{2}+
\frac{m\sigma_{0}}{2}{\dot B}^{2}+
\frac{m\sigma_{0}}{2}{\Omega_{l}}^{2}A^{2}+
\frac{3m\sigma_{0}}{2}(3{\Omega_{l}}^{2}+{\omega_{\sigma}}^{2})B^{2}+
3m\sigma_{0}{\Omega_{l}}^{2}AB
\label{hamilt}
\end{equation}
which can be done by the following transformation
\begin{equation}
\sqrt{m\sigma_{0}}A=Q_{1}\cos{\phi_{l}}-Q_{2}\sin{\phi_{l}}
\label{21}
\end{equation}
\begin{equation}
\sqrt{3m\sigma_{0}}B=Q_{1}\sin{\phi_{l}}+Q_{2}\cos{\phi_{l}}
\label{22}
\end{equation}
where the angle $\phi_{l}$ can be found from the condition of the
disappearance
of the nondiagonal term in Hamiltonian $(\ref{hamilt})$ expressed
in terms of the new coordinates $Q_{1}$ and $Q_{2}.$ The substitution
of $(\ref{21})$ and
$(\ref{22})$ into equation $(\ref{hamilt})$ gives
\begin{equation}
\cot{2\phi_{l}}=
-\frac{2{\Omega_{l}}^{2}+{\omega_{\sigma}}^{2}}
{2\sqrt{3}{\Omega_{l}}^{2}}.
\label{24a}
\end{equation}
The eigenfrequencies $\Omega_{l,j}$
are given by the following equation
\begin{equation}
({\Omega_{l}}^{2})_{1,2}=
\frac{1}{2}\Big (
4{\Omega_{l}^{2}}+{\omega_{\sigma}^{2}}\pm
\sqrt{4{\Omega_{l}^{2}}{\omega_{\sigma}^{2}}+{\omega_{\sigma}^{4}}+
16\Omega_{l}^{4}} \Big ) .
\label{16}
\end{equation}

One can clearly see that charge oscillations can be classified by
the numbers $l$ and $m$ and the index $j$ corresponding to the lower and
upper roots of equation $(\ref{16}).$
The fact that the eigenfrequencies
do not depend on the number $m$ is due to the spherical symmetry of the
considered model.


Using $(\ref{connect})$ we can find quantum mechanical
operators for amplitudes $A_{lm}$ and $B_{lm}.$ Substituting them
into the equation for classical charge density fluctuations 
we obtain the operator corresponding to this quantity
\begin{eqnarray}
\lefteqn{
\delta {\hat \sigma }(\phi, \theta)=
-\frac{e}{R^{2}}\sqrt{\frac{\hbar \sigma_{0}}{2 m}}
\sum_{lm}\sqrt{l(l+1)}Y_{lm}(\phi, \theta)   }\nonumber\\& &
\Bigg (\frac{g_{1}(\phi_{l})}{\sqrt{\Omega_{l,1}}}
({\hat {\delta}}_{lm,1}+{{\hat {\delta}}^{\dagger}}_{l,-m,1})+
\frac{g_{2}(\phi_{l})}{\sqrt{\Omega_{l,2}}}
({\hat {\delta}}_{lm,2}+{{\hat {\delta}}^{\dagger}}_{l,-m,2}) \Bigg )
\label{29a}
\end{eqnarray}
where 
${\hat {\delta}}_{lm,j}$ is the annihilation operator
of the plasmon with
numbers $lm,j.$
and the functions $g_{1}(\phi_{l})$ and $g_{2}(\phi_{l})$ are given
by the equations
\begin{equation}
g_{1}(\phi_{l})=\sqrt{3}\cos{\phi_{l}}-\sin{\phi_{l}},\
g_{2}(\phi_{l})=\cos{\phi_{l}}+\sqrt{3}\sin{\phi_{l}},
\label{30a}
\end{equation}
the angle $\phi_{l}$ is chosen to be the minimal positive
root of equation $(\ref{24a}).$

\section{Fast electron scattering}

In this section we will obtain the expression for the scattering amplitude
of fast electrons due to charge density fluctuations in ${\rm C}_{60}.$ In
the secondary quantization representation the
electron-plasmon interaction can be written as follows
\begin{equation}
{{\hat H}_{int}}=\int
{{\hat \Psi}}^{\dagger}(-e{\hat{\Phi}}){{\hat \Psi}}d{\bf r}.
\label{28}
\end{equation}
Here 
${\hat \Psi} 
=\sum_{{\bf k}} {e^{i{\bf k}{\bf r}}a_{{\bf k}}}/{\sqrt{V}}$ is the
electron field operator, ${\hat{\Phi}}$ is the
potential corresponding to the electron surface density distribution
$\delta {\hat {\sigma}}(\phi ,\theta),$
i.e. ${\hat {\Phi}}$ is the solution of the Poisson equation
with the charge density
$\delta (r-R)\delta {\hat \sigma}(\phi ,\theta).$
Hence ${\hat {\Phi}}$ and
$\delta {\hat \sigma}$
are connected by the following equation
\begin{equation}
{\hat \Phi}=\int \frac{\delta (r^{\prime}-R)
\delta {\hat {\sigma}} (\phi^{\prime},\theta^{\prime})}
{|{\bf r}-{\bf r}^{\prime}|}d{\bf r}^{\prime}.
\label{29}
\end{equation}

Substituting equation $(\ref{29})$ and the explicit
expressions for the electron field and the electron density
fluctuation operators (see (\ref{29a}) )
into equaiton $(\ref{28})$
we obtain the expression for the electron-plasmon
interaction
\begin{eqnarray}
\lefteqn{
{{\hat H}_{int}}=-\sum_{{\bf k},{\bf k}^{\prime}}
{{a^{\dagger}}_{{\bf k}^{\prime}}}
{{a}_{\bf k}}
\frac{{(4\pi)}^{2}e^{2}}{{|{\bf k}-{{\bf k}^{\prime}}|}^{2}V}
\sqrt{\frac{\hbar \sigma_{0}}{2m}}      }\nonumber\\& &
\sum_{lm,j}
i^{l}\sqrt{\frac{l(l+1)}{\Omega_{l,j}}}
g_{j}(\phi_{l})j_{l}(|{\bf k}-{{\bf k}^{\prime}}|R)
Y_{lm}\bigg (\frac{{{\bf k}^{\prime}}-{\bf k}}{|{{\bf k}^{\prime}}-{\bf k}|}
\bigg )
({{\hat \delta}_{lm}}+{{{\hat \delta}^{\dagger}}_{l,-m}} ).
\end{eqnarray}
We shall consider only the case of fast scattering electrons, i.e.
the only relevant processes are
those in connection with the creation of one plasmon.
Hence we can use the Born approximation. The quantitative criterium
of the applicability of the Born approximation will be discussed further.

In the initial state all the plasmon occupation numbers are equal to zero.
Let us suppose that
after scattering only the occupation number of the state $lm,j$ is
equal to one.
Defining the initial and final momenta of the scattering electron
${\bf k}$ and ${{\bf k}^{\prime}}$ respectively we obtain the
following expression for
the modulus squared of the matrix element of ${{\hat H}_{int}}$
\begin{equation}
|\langle {\bf k};0|{{\hat H}_{int}}|{{\bf k}^{\prime}};lm,j\rangle |^{2}=
\frac{(4\pi)^{4}e^{4}\hbar \sigma_{0}}{2m{f}^{4}V^{2}}
\frac{l(l+1)}{{\Omega}_{l,j}}
{g_{j}}^{2}(\phi_{l})
{j_{l}}^{2}(q R)
Y_{lm}\left(\frac{{\bf q}}{q}\right)
Y_{lm}^{*}\left(\frac{{\bf q}}{q}\right)
\end{equation}
where ${{\bf q}}={{\bf k}^{\prime}}-{\bf k}.$

As each eigenfrequency $\Omega_{l,j}$ is $2l+1$-fold degenerated,
we shall calculate the total cross-section for the transferred energy
$\hbar \Omega_{l,j}.$ Summing up the contributions of plasmons
with given $l$ and $j$ over $m$
we get
\begin{equation}
\frac{d\sigma_{l,j}}{d\Omega_{{{\bf} k}^{\prime}}}=
\int\big (
\sum_{m}
{|\langle {\bf k};0|{{\hat H}_{int}}|{{\bf k}^{\prime}};lm,j\rangle |}^{2}
\big )
\delta (\omega_{{\bf k}^{\prime}}+{\Omega}_{l}-{\omega}_{\bf k})
\frac{2\pi V}{{\hbar}^{2}v}
\frac{{k^{\prime}}^{2}d{k^{\prime}}V}{{(2\pi)}^{3}}.
\label{31}
\end{equation}
In the above equation $\omega_{{\bf k}}={\hbar k^{2}}/{2m},$
$v$ is the initial velocity of the electron.

Using the equation
\begin{equation}
\sum_{lm}Y_{lm}({\bf q}/{q})
Y_{lm}^{*}({\bf q}/{q})=
{(2l+1)}/{4\pi}
\end{equation}
( see Ref.\cite{21}) we get after some algebra
\begin{equation}
\frac{d\sigma_{l,j}}{d\Omega_{{\bf k}^{\prime}}}=
8\pi\frac{{e^{4}}{\sigma_{0}}}{{\hbar}^{2}v}
l(l+1)(2l+1)\frac{{g_{j}}^{2}(\phi_{l})}{\Omega_{l,j}}
\sqrt{k^{2}-\frac{2m{\Omega_{l,j}}}{\hbar}}
\frac
{j_{l}^{2}(|{{\bf k}^{\prime}}-{\bf k}| R)}
{{|{{\bf k}^{\prime}}-{\bf k}|}^{4}}
\label{32}
\end{equation}

  \section{Results and Discussion}

In experiments one can measure the number of electrons
reaching a detector, i.e. one measures the function $(\ref{32})$
integrated over a certain solid angle. Let this solid angle
be restricted by two cones of angles ${\theta}_{1}$ and
${\theta}_{2}$ respectively. In this case we can write for the total
cross-section $\sigma_{l,j} ({\theta}_{1},{\theta}_{2})$
\begin{equation}
\sigma_{l,j} (\theta_{1},\theta_{2})=
\frac{2\pi e^{4}N}{\hbar {\Omega}_{l,j} E}l(l+1)(2l+1)
g_{j}^{2}(\phi_{l})
\int
\limits
_{|{\bf k}-{\bf k}_{1}^{\prime}|R}
^{|{\bf k}-{\bf k}_{2}^{\prime}|R}
\frac{j_{l}^{2}(x)}{x^{3}}dx
\label{33a}
\end{equation}
where $E={m v^{2}}/{2},$ $|{\bf k}-{\bf k}_{i}^{\prime}|=
{\sqrt{k^{2}+{{k}^{\prime}}^{2}-2k{{k}^{\prime}}\cos{\theta_{i}}}},
\ i=1,2.$

We shall consider the case of small transferred energy, i.e.
${\omega}_{{\bf k}}-{\omega}_{{\bf k}^{\prime}}\ll {\omega}_{{\bf k}}.$
In principle, several different situations are possible. If both the
angles $\theta_{1}$ and $\theta_{2}$ are small, the integration limits
in the integral in $(\ref{33a})$ are also small and we can expand the spherical
Bessel function into a Taylor series and take only the first term of the
expansion. As $j_{l}(x)\cong{x^{l}}/{(2l+1)!!}$ if $x\rightarrow 0 ,$
we get
\begin{equation}
\sigma_{l,j}(\theta_{1},\theta_{2})\sim
\int
\limits
_{|{\bf k}-{\bf k}_{1}^{\prime}|R}
^{|{\bf k}-{\bf k}_{2}^{\prime}|R}
x^{2l-3}dx
\end{equation}
and one can clearly see that the contribution
of the plasmons with $l=1$ is dominant.
In the opposite situation
${|{\bf k}-{\bf k}_{i}^{\prime}|R}\gg 1,\ i=1,2$
we can use the asymptotic expression for the spherical Bessel function
$j_{l}(x)\cong{\sin{(x-{\pi l}/{2})}}/{x}$ if $x\rightarrow\infty .$
As the integration limits do not depend on $l$ significantly, we can see that
the values of the integral in $(\ref{33a})$ are of the same order for different
$l.$ Hence the relative contributions of plasmons with different $l$
depend only on the factor ${g_{l,j}^{2}(\phi_{l})}/{\Omega_{l,j}}.$
In the case
${|{\bf k}-{\bf k}_{i}^{\prime}|R}\sim 1,\ i=1,2$
it is possible to estimate the integral in $(\ref{33a})$ by the substitution
$j_{l}(x)\sim {x^{l}}/{(2l+1)!!}.$ One can clearly see that only the
contribution of the plasmons with small $l$ is relevant because of the rapidly
increasing multiplier ${\big ( (2l+1)!! \big )}^{2}$ in the denominator.

Strictly speaking the above analysis is not applicable for the plasmons with
$l>3.$ The ${\rm C}_{60}$ symmetry group is icosahedral and it does not have
irreducible representations with a dimension
of more than five. Hence the plasmons
with $l>3$ in the spherical model split \cite{22,23}
and the classification of plasmons
in the real ${\rm C}_{60}$ molecule is different. We will not discuss further 
this difference because in the experiments which we will analyse only 
the contribution of the plasmons $l=1,2$ is relevant. On the other hand, 
the small size of ${\rm C}_{60}$ leads to the strong damping of plasmons.
The estimation for the spherical model in Ref.\cite{16} shows that
the plasmons with $l>3$ could hardly exist.

We shall compare our results to the experiment from Ref.\cite{5} .
In this work the number of electrons with $E=1\ {\rm keV}$
scattering in the solid angle ${0.5}^{\circ}<\theta <{2.5}^{\circ}$
(i.e. $\theta_{1}={0.5}^{\circ},\ \theta_{2}={2.5}^{\circ}$) per
unit time was measured. The average cross-section
\begin{equation}
{\tilde \sigma}_{l,j}(\theta_{1},\theta_{2})=
{\sigma_{l,j}}(\theta_{1},\theta_{2})
/{2\pi (\cos{\theta_{1}}-\cos{\theta_{2}})}
\end{equation}
obtained 
using the spherical model for the plasmons with $l=1\dots 5$
is shown in Table~\ref{table}. We used the value 
$\omega_{\sigma}=16\ {\rm eV}$
for the characteristic $\sigma$-electron transition
frequencies and 
$R=0.4{\rm \ nm}$
for the radius of the sphere in the model.
For the conditions of the analised experiments
${|{\bf k}-{\bf k}_{i}^{\prime}|R}\sim 1,\ i=1,2$
and one can clearly see that the contribution of the plasmons with
$l>2$ decays rapidly. This is in good agreement with the qualitative
consideration.
The $l=1$ $\pi$- and $\sigma$-plasmons and the quadrupole $\sigma$-plasmon
make the most significant contribution to the cross-section.

Plasmons 
in $C_{60}$ are characterized by strong damping
due to the small size of the molecule: in small objects the energy
of plasma oscillations per one electron ${{\hbar}{\Omega}}/{N}$
is strongly enhanced and leads to strong electron-electron scattering, 
meanwhile for bulk plasmons this energy is negligible.
As energies of $\pi$- and $\sigma$-plasmons do not depend on 
$l$ significantly 
(see Table\thinspace\ref{table}), the parameter 
${\hbar\Omega}/{N}$ is approximately the same for $\pi$- and $\sigma$-plasmons
respectively and we can expect that the damping of plasmons 
does not depend on $l$ strongly.

In the above calculation of the cross-section we did not take into account
plasmon damping. 
The characteristic inverse life time 
$\eta_{j}$ taken from the
experiment is equal to $1.5$ and $8\  {\rm eV}$ for ${\pi}$- and
$\sigma$-plasmons respectively. The function
\begin{equation}
\sigma (\omega )=\sum_{l,j}
\frac{{\tilde \sigma}_{l,j}(\theta_{1},\theta_{2}){\eta_{j}}}{\pi}
\frac{1}{{(\omega-\Omega_{l,j})}^{2}+{\eta_{j}}^{2}}
\end{equation}
is plotted in 
Fig.~1
together with the experimental curve. One 
can see that our results are in good agreement with the experiment.

As it can be seen from Table\thinspace\ref{table} and 
Fig.~1
the first peak on the graph corresponds to the dipole $\pi$-plasmon;
the second peak corresponds to the dipole and quadrupole $\sigma$-plasmons.
The contributions of the other plasmons are small.
It has already been discussed 
that ratios ${\sigma_{l,1}}/{\sigma_{l,2}}$
are proportional in our case to ratios 
$(g_{l,1}^{2}/{\Omega_{l,1}})/(g_{l,2}^{2}/{\Omega_{l,2}}).$
Note that although $\Omega_{l,1}<\Omega_{l,2}$ we have for the cross-sections
${\sigma_{l,1}}>{\sigma_{l,2}}$ and the ratio 
${\sigma_{l,1}}/{\sigma_{l,2}}$ decreases if $l$ increases.
This is the consequence of the fact that $g_{l,1}\rightarrow 0$
if $l\rightarrow\infty :$  
for large $l$ we can write
using equation $(\ref{24a})$ 
$\cot{\phi_{l}}\rightarrow -{1}/{\sqrt{3}},$ hence ${g_{l,1}}\rightarrow 0$
(see $(\ref{30a})$ ) and we obtain an interesting result:
the ratio ${\sigma_{l,1}}/{\sigma_{l,2}}$ tends to zero if
$l\rightarrow\infty .$

Let us discuss the condition of the applicability of the Born approximation.
The Born approximation is valid if 
$|U|a\ll {\hbar v}$ (see Ref.\cite{21}) where $|U|$ and $a$ are 
the characteristic value of the potential
and its radius respectively, $v $ is the velocity of the 
scattering particle. In our case the characteristic value of 
charge density fluctuations $\delta \sigma$ is 
${({e}/{a^{2}})}\sqrt{ {\hbar \sigma_{0}}/{m \Omega}}$ 
where $\Omega\sim 10\ {\rm eV}$ is the characteristic oscillation frequency.
Hence $|U|\sim {({e^{2}}/{a})}\sqrt{{\hbar \sigma_{0}}/{m\Omega}}$
and the condition of the applicability can be written as follows
\begin{equation}
e^{2}\sqrt{\frac{\hbar \sigma_{0}}{m\Omega}}\ll \hbar v .
\label{33}
\end{equation}

Substituting 
$\sigma_{0}= {N}/{4\pi R^{2}},$ $a\sim R,$
$R=0.4\ {\rm nm}$
into $(\ref{33})$ we see
that the Born approximation is applicable if the velocity
of the scattering electron satisfies the condition
$v\gg 2\cdot {10}^{8}$\ cm/s. In the analysed experiments
the electron velocity is $1.88\cdot {10}^{9}$\ cm/s, i.e.
the Born approximation is applicable.

Briefly summarizing we have shown that the simple two-fluid model
used by the authors of \cite{16} for the predictions 
of the collective spectrum of ${\rm C}_{60}$ turned out to be
good for the quantitative calculations of the intensities of the plasmon
peaks in EELS spectra. The advantage of the description based on this
model is that it is physically and technically simple and the model itself
does not include any fitting parameters (with the exception of the decay
rates which however can not be properly calculated in more
sophisticated techniques\cite{5}). All this makes the model useful
for the quantative consideration of the collective behaviour of electrons in
more complicated fullerene systems.

\begin{table}
\caption{
Frequencies and intencities of plasmon peaks 
calculated using formulae (17) and (27)
for 1keV incindent electrons at  
${1.5}^{\circ}\pm {1}^{\circ}$.}     
\begin{tabular}{ccc}
Type of plasmon&Plasmon frequency, eV&Cross section, ${\rm nm}^{2}/{\rm sr}$\\
\tableline
l=1,\ j=1&5.95&3.82\\
l=1,\ j=2&22.21&9.56\\
l=2,\ j=1&6.68&0.69\\
l=2,\ j=2&26.49&5.16\\
l=3,\ j=1&7.03&0.09\\
l=3,\ j=2&30.12&1.35\\
l=4,\ j=1&7.23&$8.4\cdot 10^{-3}$\\
l=4,\ j=2&33.35&0.22\\
l=5,\ j=1&7.36&$5.6\cdot 10^{-4}$\\
l=5,\ j=2&36.29&0.024\\
\end{tabular}
\label{table}
\end{table}
\vskip1.5cm

\centerline{\epsfxsize=10cm \epsfbox{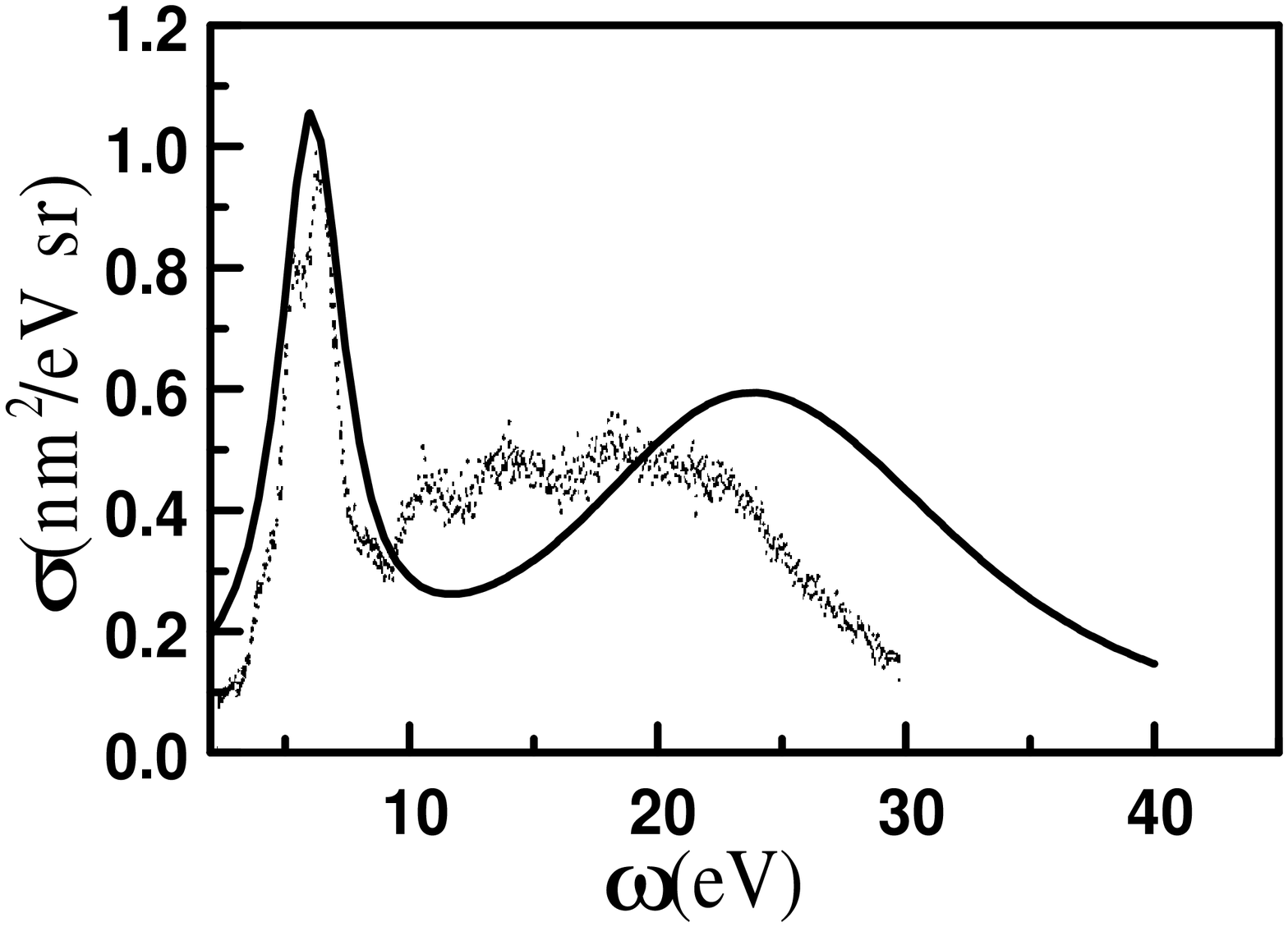}}
{
\vskip0cm
\footnotesize {\bf Fig.1}~Measured in 
${\rm Ref.}^{5}$
and calculated using equation (27)
(solid line) inelastic cross-section with incident 1 keV electrons
at 
${1.5}^{\circ}\pm {1}^{\circ}$.}
\vskip0.3cm
\hskip-0.7cm

This work was fulfilled within Russian research and development programm
``Fullerenes and Atomic Clusters'' project N 94014.




%
%

\end{document}